\documentclass[a4paper,9pt]{article}

\usepackage[a4paper]{geometry}
\usepackage{amsmath}
\usepackage[T1]{fontenc}
\usepackage{amsmath}
\usepackage{amssymb}
\usepackage{textcomp}
\usepackage{amsthm}
\usepackage{amsfonts}
\usepackage{mathrsfs}
\usepackage{array}
\usepackage[english]{babel}
\usepackage{lscape}
\usepackage{graphicx}
\usepackage[all]{xy}

\newtheorem{theorem}{Theorem}[section]
\newtheorem{definition}{Definition}[section]

\numberwithin{equation}{section}
\title{Vectorial Feedback with Carry Registers and Memory requirements}
\author{Abdelaziz MARJANE, Abdellah MOKRANE and Boufeldja ALLAILOU\\
LAGA, UMR CNRS 7539, Universit\'e Paris 13, Villetaneuse, France\\
LAGA, UMR CNRS 7539, Universit\'e Paris 8, Saint-Denis, France \\
LAGA, UMR CNRS 7539, Universit\'e Paris 8, Saint-Denis, France    \\ {marjane, allailou, mokrane@math.univ-paris13.fr}
}

\begin{document}
\maketitle

\begin{abstract}
In \cite{marjane2010}, we have introduced vectorial
conception of FCSR's in Fibonacci mode. This conception allows us to easily analyze FCSR's
over binary finite fields $\mathbb{F}_{2^{n}}$ for $n\geq 2$. In \cite{allailou2010}, we describe and study the corresponding Galois
mode and use it to design a new stream cipher. In this paper, we introduce the Ring mode for vectorial FCSR, explain the analysis of such Feedback registers and illustrate with a simple example.\\
keywords:LFSR, FCSR, stream ciphers, 2-adic, sequences, Vectorial register\end{abstract}

\maketitle

\section{Introduction} 
The Ring mode was first introduced for LFSR's in \cite{RingLFSR} and adapted
to binary FCSR in \cite{RingFCSR}. In this mode, any cell can be used as a
feedback bit for any other cell.  Registers in Ring
mode are represented by a matrix which can be chosen arbitrarily. 
The classical Fibonacci and Galois modes are in fact
special cases of the Ring mode. Recall the notion of LFR and Ring mode.
\begin{definition}[LFR]Let $n$ and $r$ be two positive integers and $T$ a square $r\times r$ matrix with coefficients in the binary field $\mathbb{F}_{2^{n}}$. A Linear Feedback Register (LFR) over $\mathbb{F}_{2^{n}}$ of length $r$ with  transition matrix $T$ is a sequence generator whose state is an element $s(t)=(a_{0}(t),\ldots,a_{r-1}(t))\in (\mathbb{F}_{2^{n}})^{r}$ and whose operation state change is given by $s(t+1)=s(t).T$. 
\end{definition}

The Ring mode corresponds to the case where the matrix $T=(t_{i,j})_{i,j}$ is such that $t_{i+1,i}=1$ and $t_{1,r}\neq 0$.
This mode generalizes both Fibonacci and Galois modes given respectively by the following transition matrix :
\begin{equation}\label{transition}
F=\left(\begin{array}{ccccc}
0&\ldots&0&q_{r}\\
1&\ldots&0&q_{r-1}\\
\vdots&\ddots&\vdots&\vdots\\
0&\ldots&1&q_{1}
\end{array}\right)\,G=\left(\begin{array}{ccccc}q_{1}&\ldots&q_{r-1}&q_{r}\\
1&\ldots&0&0\\
\vdots&\ddots&\vdots&\vdots\\
0&\ldots&1&0
\end{array}\right).
\end{equation}
\begin{theorem}[\cite{goresky2009},p.268]The output sequence of an LFR with transition matrix $T$ can be generated by an LFSR 
with connection polynomial equal to $\det(I-XT)$.
\end{theorem}
\noindent
So from a theoritical point of view, LFRs are no more powerful than LFSRs but they 
can provide efficient software implementations by reducing the number of connections and operations (see \cite{goresky2009}). FCSR is a class of non linear FSR with good properties as for LFSR. In this paper, after review of different modes of binary FCSR and vectorial FCSR, we introduce the analog of LFR for registers with carry over $\mathbb{F}_{2^{n}}$ in a general setting and establishe its basic properties. To be more precise, fix a primitive polynomial $P(X)$ of degree $n$ over $\mathbb{F}_{2}$ and $T$ a square $r\times r$ matrix with coefficients in the binary field $\mathbb{F}_{2^{n}}\cong \mathbb{F}_{2}[X]/(P(X))$. We associate to $T$ in a canonical way a $nr\times nr$ square matrix $\mathcal{T}$ with coefficients in $\mathbb{Z}$ and define Feedback with carry registers over $\mathbb{F}_{2^{n}}$ of length $r$ with  transition matrix $T$ as a sequence generator whose state is an element pair $(a(t),m(t))$ where $a(t)=(a_{0}(t),\ldots,a_{r-1}(t))\in (\mathbb{F}_{2^{n}})^{r}$ and $m(t)=(m_{1}(t),\ldots,m_{r}(t))\in (\mathbb{Z}^{n})^{r}$ and whose operation state change is given by 
$$\begin{array}{rcl}
a(t+1)=\Big(a(t)\otimes\mathcal{T}\oplus m(t)\Big)\mathtt{mod}2\\
m(t+1)=\Big(a(t)\otimes\mathcal{T}\oplus m(t)\Big)\mathtt{div}2
\end{array}$$
where $\otimes$ is defined in section 5. We prove the following structural theorem:
\begin{theorem}The $2$-adic expansion $\sum\limits_{t=0}^{t=+\infty}c(t)2^{t}$ where $c(t)$ is any binary component of $a_{i}(t)$ is equal to a rational number $\frac{p}{q}$ where $q=\det(I_{rn}-2\mathcal{T})$.
\end{theorem}
\section{Binary Feedback with Carry Registers in Differents Modes}
Feedback with carry shift registers or FCSRs were developped by Goresky and Klapper  \cite{goresky1994} \cite{goresky1997} and \cite{Galois2002}. These registers rely over a $2$-adic elegant structure which is an alternative to the linear architecture of LFSRs. They differ from LFSRs by adding memories and using computations over $\mathbb{Z}$. 
\begin{definition} A binary FCSR in Fibonacci mode of length $r$ and connection coefficients $q_{1},\ldots,q_{r}\in \left\{0,1\right\}$ is an automaton sequence generator whose state is an element $(a_{0},a\ldots,a_{r-1},m_{r-1})$ where $a_{i}\in\{0,1\}$ for all $i$ and $m_{r-1}\in\mathbb{Z}$ and whose operation state change is given by the following procedure: 
\begin{itemize}
\item  Compute the integer $\sigma_{r}=q_{r}a_{0}+\ldots+q_{1}a_{r-1}+m_{r-1}$ in $\mathbb{Z}$. 
\item Compute $a_{r}=\sigma_{r}\pmod 2$ and $m_{r-1}=\sigma_{r} \mathtt{div}\, 2$. 
\item Output $a_{0}$ and $m_{r-1}$, shift the other coefficients $a_{1},\ldots,a_{r-1}$ and enter $a_{r}$ and $m_{r}$. 
\end{itemize}
$(a_{0},a_{1},\ldots)$ is called the output sequence and $q=q_{r}2^{r}+\ldots+q_{1}2-1$ is called the connection integer of the FCSR.
\end{definition}
\begin{definition}A binary FCSR in Galois mode of length $r$ with connection coefficients $q_{1},\ldots ,q_{r}\in \left\{0,1\right\}$ is an automaton whose state at the $t$ th steps is an element $s(t)=(a_{0}(t),\ldots ,a_{r-1}(t),m_{1}(t),\ldots,m_{r}(t))\in\left\{0,1\right\}^{r}\times \mathbb{Z}^{r}$ and whose state change operation is as follows: 
\begin{itemize}
\item Compute $\sigma_{i}(t+1)=q_{i}a_{0}(t)+a_{i+1}(t)+m_{i+1}(t)$ for all $0\leq i\leq r-2$ and $\sigma_{r-1}(t+1)=q_{r}a_{0}(t)+m_{r}(t)$.
\item Compute $a_{i}(t+1)=\sigma_{i}(t+1)\pmod 2$ and $m_{i+1}(t+1)=\sigma_{i}(t+1)\mathtt{div}2$ for all $1\leq i\leq r$. 
\item Output $a_{0}(t)$ and replace $a_{i}(t)$ by $a_{i}(t+1)$ and $m_{i+1}(t)$ by $m_{i+1}(t+1)$ for all $1\leq i\leq r$. 
\end{itemize}
 $s(0)$ is the initial state, $(a_{0}(0),a_{0}(1),a_{0}(2),\ldots)$ the output sequence.
\end{definition}
Unlike the Fibonacci mode, all cells are simultaneously updated in Galois mode. Galois mode is more convenient for cryptographic applications. Whatever the mode, we associate a $2$-adic integer $\sum\limits_{i=0}^{i=+\infty}a_{i}2^{i}$ to the output sequence.
\begin{theorem} The 2-adic integer associated to the output sequence is a rational $\frac{p}{q}$ where $q$ is the connection integer (Definition 2), $$\begin{array}{rl}-p=\sum\limits_{i=0}^{i=r-1}a_{i}2^{i}+m_{r-1}2^{r}-\sum\limits_{i=1}^{k=r-1} \sum\limits_{j=1}^{j=i}q_{i}a_{i-j}2^{i}&\textrm{ in Fibonacci mode and}\\ -p=\sum\limits_{i=0}^{i=r-1}a_{i}(0)2^{i}+\sum\limits_{i=1}^{i=r}m_{i}(0)2^{i}&\textrm{ in Galois mode}.
\end{array}$$
\end{theorem}

FCSR sequences have good randomness properties like periodicity, distribution of block, balanced property, maximal period sequences called $l$-sequences, cross-correlation of two level, etc.

The Ring mode for FCSR developped in \cite{RingFCSR} generalizes both Fibonacci and Galois modes and has many advantages over these both modes.
\begin{definition}[FCR]A binary Feedback with Carry Register (FCR) of length $r$ with transition matrix $T$ is a sequence generator whose state is a pair $(a(t),m(t))$ where $a(t)=(a_{0}(t),\ldots,a_{r-1}(t))\in\{0,1\}^r$ and $m(t)=(m_{1}(t),\ldots,m_{r}(t))\in \mathbb{Z}^r$; and whose operation state change is given by 
\begin{equation}
\label{E7}
a(t+1))=\Big(a(t).T+m(t)\Big)\mathtt{mod} 2\textrm{ and }m(t+1)=\Big(a(t).T+m(t)\Big)\mathtt{div}2.
\end{equation}
\end{definition}
Fibonacci and Galois modes of FCSR can be represented as a Ring FCSR with a special transition matrix of the form (\ref{transition}). The analysis of binary FCR can be made as in the Fibonacci case. 
\begin{theorem} The output sequence $(a_{i}(0),a_{i}(1),\ldots)$ of a binary FCR defines a 2-adic integer which is a rational number $\frac{p_{i}}{q}$ where $q=\det(I-2T)$.
\end{theorem}
To generate $l$-sequences in Ring mode, we have to choose a matrix $T$ such that $\det(I-2T)$ is prime and 2 is a primitive root modulo $\det(I-2T)$. Unfortunately there is no simple method for general $T$ to do this.

\section{Vectorial FCSR in Fibonacci mode}
\begin{figure}[!h] 
\begin{center}
\includegraphics[height=10cm,width=11cm]{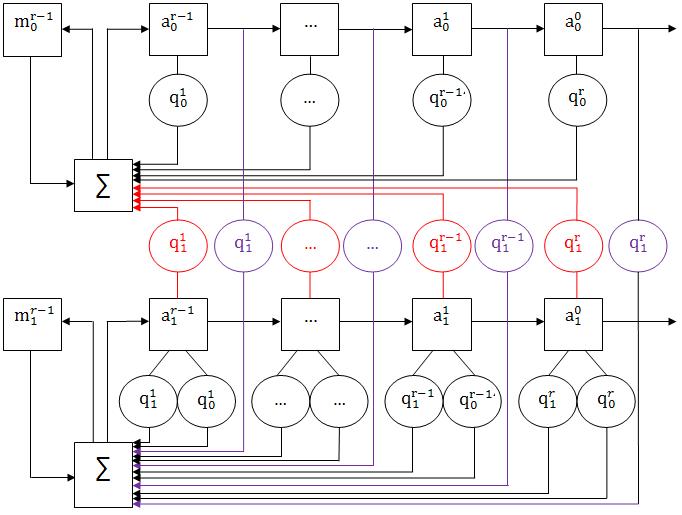}
\end{center}
\caption{VFCSR-Q in Fibonacci mode.}
\label{VFCSR-Q}
\end{figure}

To construct FCSR over any finite fields $\mathbb{F}_{2^{n}}$, we use a vectorial conception introduced by Klapper \cite{klapper1994}. We have completely developed the vectorial analysis of these registers \cite{marjane2010}. They present the same basic properties as in the binary case. 
\paragraph{\textbf{Description of the Automaton:}} Let $P$ be a primitive polynomial over $\mathbb{F}_{2}$ of degree $n$. $\mathbb{F}_{2}[X]/(P)$ is a vector space of dimension $n$ over $\mathbb{F}_{2}$, we consider its canonical basis $\left\{1,\bar{X},\ldots,\bar{X}^{n-1}\right\}$. $P$ is identified to its canonical lift in $\mathbb{Z}[X]$ and consider the free $\mathbb{Z}$-module $\mathbb{Z}[X]/(P)$ of rank $n$ and its canonical basis $\mathcal{B}=\left\{1,\bar{X},\ldots,\bar{X}^{n-1}\right\}$. 
\begin{definition} A Vectorial FCSR in Fibonacci mode over $(\mathbb{F}_{2},P,\mathcal{B})$ of length $r$ with connection coefficients $q_{1},\ldots,q_{r}\in \mathbb{F}_{2}[X]/(P)$ is an automaton whose state is an element $s=(a_{0},\ldots,a_{r-1},m_{r-1})$ where $a_{i}\in\mathbb{F}_{2}[X]/(P)$ and $m_{r-1}\in\mathbb{Z}[X]/(P)$ and whose state change operation is described as follows: 
\begin{itemize}
\item Express the elements $a_i,q_i, m_i$ in the basis $\left\{1,\bar{X},\ldots,\bar{X}^{n-1}\right\}$. 
$$\begin{array}{rcl}
\forall i\in \mathbb{N},&a_{i}=a_{0}^{i}+a_{1}^{i}\bar{X}+\ldots+a_{n-1}^{i}\bar{X}^{n-1}&\textrm{ where }a_{j}^{i}\in \left\{0,1\right\},\\
\forall 1\leq i\leq r,&q_{i}=q_{0}^{i}+q_{1}^{i}\bar{X}+\ldots+q_{n-1}^{i}\bar{X}^{n-1}&\textrm{ where }q_{j}^{i}\in \left\{0,1\right\},\\
\forall i \geq r-1,&m_{i}=m_{0}^{i}+m_{1}^{i}\bar{X}+\ldots+m_{n-1}^{i}\bar{X}^{n-1}&\textrm{ where }m_{j}^{i}\in \mathbb{Z}.
\end{array}$$
\item Take the canonical lift of $a_{i}$ and $q_{i}$ in $\mathbb{Z}[X]/(P)$ with respect $\mathcal{B}$. 
\item Compute $\sigma_{r}=q_{r}a_{0}+\ldots+q_{1}a_{r-1}+m_{r-1}$ as a vector in $\mathcal{B}$. 
\item Compute the coordinates of $a_{r}$ and $m_{r}$ with respect $\mathcal{B}$: 
\begin{equation}\label{E2}
a_{j}^{r}=\sigma_{j}^{r} \pmod 2\textrm{ and }
m_{j}^{r}=\sigma_{j}^{r}(\mathtt{div}2)=\frac{1}{2}(\sigma_{j}^{r}-a_{j}^{r}).
\end{equation}
\end{itemize}

The feedback function is $f(s)=(a_{1},\ldots,a_{r},m_{r-1})$ and the output function is $g(x_{0},\ldots,x_{r-1},y)=x_{0}$. The VFCSR generate a vectorial sequence\\ $\underline{a}=(g(s),g(f(s)),g(f^{2}(s)),\ldots)=(a_{0},a_{1},a_{2},\ldots)$.
\end{definition} 
Figure \ref{VFCSR-Q} illustrates a VFCSR over $(\mathbb{F}_{2},X^{2}-X-1,\mathcal{B})$ called VFCSR-Q in Fibonacci mode.
\paragraph{\textbf{Analysis:}} We decompose the output sequence $\underline{a}$ into $n$ components $\underline{a}_{j}=(a_{j}^{0} ,a_{j}^{1}, \cdots)$ and associate to each component its 2-adic expansion $\beta_{j}=a_{j}^{0}+a_{j}^{1}2+\cdots$ and form a 2-adic vector $\beta=(\beta_{j})_{j}$. The connection integer $q=q_{r}2^{r}+\ldots+q_{1}2-1$ is an element in $\mathbb{Z}[X]/(P)$ and its components with respect $\mathcal{B}$ are $(\tilde{q}_{0}-1,\tilde{q}_{1},\ldots,\tilde{q}_{r})$ where $\tilde{q}_{j}=q^{r}_{j}2^{r}+\ldots+q^{1}_{j}2$. We call $(\tilde{q}_{0},\ldots,\tilde{q}_{r})$ \textit{the connection vector of the VFCSR}. Using simple computations, we show that $\beta$ is a solution of a linear system with integral coefficients represented by an invertible $n\times n$ matrix called \textit{the connection matrix of the VFCSR} and denoted $\mathcal{M}$. Note that there is a subtile relation between the transition matrix $T$ used in the conception of a binary Ring mode and the connection matrix $\mathcal{M}$ used in the analysis of a Vectorial FCSR (see Example 1 after Theorem 7).
\begin{theorem}Consider a VFCSR in Fibonacci mode over $(\mathbb{F}_{2},P,\mathcal{B})$ of length $r$ with connection vector $(\tilde{q}_{0},\ldots ,\tilde{q}_{n-1})$, connection integer $q$ and connection matrix $\mathcal{M}$. Then for any sequence $\underline{a}$ generated by this VFCSR, the associated 2-adic vector $\beta $ is in $\frac{1}{|\det \mathcal{M}|}\mathbb{Z}^{n}$ and $|\det \mathcal{M}|$ is odd. $\mathcal{M}$ is the matrix in the canonical basis $\mathcal{B}$ of the linear  transformation defined as the multiplication by $-q$ and
$\det(\mathcal{M})=\mathtt{N}(-q)=(-1)^{n}\mathtt{N}(q)$ where $\mathtt{N}=\mathtt{N}_{\mathbb{Q}}^{\mathbb{Q}[X]/(P)}$ is the norm of the number field $\mathbb{Q}[X]/(P)$ over $\mathbb{Q}$.
\end{theorem}
The components sequences $\underline{a}_{j}$ are all periodic and the periods divide the order of 2 modulo $|\mathtt{N}(q)|$. The period of $\underline{a}$ is the $\mathtt{lcm}$ of the periods of the components sequences. We denote $|\mathtt{N}(q)|$ by $\tilde{q}$ and call it \textit{the connection norm of the VFCSR}. $\tilde{q}$ can be represented as an $n$-form with arguments $(\tilde{q}_{0},\ldots,\tilde{q}_{r-1})$. This $n$-form is determined by the form of the connection matrix. To generate sequences with maximal period, we must generate numbers $\tilde{q}$ such that $\tilde{q}$ is a prime, $2$ is a primitive root modulo $\tilde{q}$ and $\tilde{q}$ is represented by the $n$-form defined by $\mathcal{M}$. For example, in the case where $n=2$, $\tilde{q}$ must be represented by the quadratic form $u^{2}+uv-v^{2}$ with $u=\tilde{q}_{0}-1$ and $v=\tilde{q}_{1}$.
\paragraph{\textbf{Pseudorandom Properties of VFCSRs:}}
VFCSRs sequences have good pseudorandom properties. In fact, we have tested VFCSR in the quadratic case $(n=2)$ for several triplets $(\tilde{q},u,v)$ given in Table \ref{triplet}, using the package NIST STS  \cite{marjane2010}. This package consists of 15 different statistical tests like perfect balance, good uniform distribution, the Matrix rank, the Maurer test, the compressibility of sequences, etc\ldots For the quadratic case, we have two components sequences $\underline{a}_{0}$ and $\underline{a}_{1}$ which have passed succesful all statistical tests. To read Table \ref{triplet}, $l_{x}$ is the 2-adic length of $x$ and is the size of the corresponding binary FCSR; and $l_{(x,y)}=\max(l_{x},l_{y})$ is the size of the corresponding VFCSR-Q.
\begin{table}[!h]
\begin{center}
\begin{tabular}{|l|l|l|l|l||l|l|l|l|l|}
\hline
$l_{\widetilde{q}}$ & $\widetilde{q}$ & $l_{(u,v)}$ & $u$ & $v$ & $l_{\widetilde{q}}$ & $\widetilde{q}$ & $l_{(u,v)}$ & $u$ & $v$  \\ 
\hline
4 & 11 & 2 & 3 & 2 & 16 & 101419 & 8 & 331 & 354\\ 
\hline
4 & 11 & 5 & 31 & 50 & 16 & 109891 & 8 & 331 & 330\\ 
\hline
10 & 1259 & 5 & 35 & 34 & 16 & 115259 & 8 & 339 & 338\\ 
\hline
9 & 829 & 5 & 35 & 44 & 16 & 103451 & 8 & 339 & 370 \\ 
\hline
13 & 8821 & 6 & 85 & 28 & 16 & 112181 & 8 & 351 & 380\\ 
\hline
11 & 2389 & 6 & 85 & 124 & 16 & 121421 & 8 & 351 & 332\\ 
\hline
12 & 8179 & 6 & 89 & 86 & 17 & 132499 & 8 & 373 & 390\\ 
\hline
11 & 3581 & 6 & 89 & 124 & 17 & 157141 & 8 & 373 & 316\\ 
\hline
13 & 9949 & 6 & 95 & 84 & 18 & 389219 & 9 & 637 & 662\\ 
\hline
12 & 7621 & 6 & 95 & 108 & 18 & 395429 & 9 & 651 & 692 \\ 
\hline
18 & 411491 & 9 & 639 & 
634 &&&&&\\ \hline
 18 & 424451 & 9 & 651 & 
650 &&&&&\\ \hline
 18 & 428339 & 9 & 657
& 662 &&&&&\\ \hline
 18 & 443771 & 9 & 657
& 638 &&&&&\\ \hline
 18 & 467171 & 9 & 683
& 682 &&&&&\\ \hline
 18 & 481619 & 9 & 
675 & 634 &&&&&\\ \hline
 18 & 502499 & 9 & 689
& 646 &&&&&\\ \hline
 20 & 1164589 & 9 & 
1001 & 204 &&&&&\\ \hline
 20 & 3932741 & 10 & 
2001 & 2036 &&&&&\\ \hline
\end{tabular}
\end{center}
\caption{Some triplets and their length.}
\label{triplet}
\end{table}
\section{Vectorial FCSR in Galois mode }
\begin{figure}[!h] 
\begin{center}
\includegraphics[height=11cm,width=11cm]{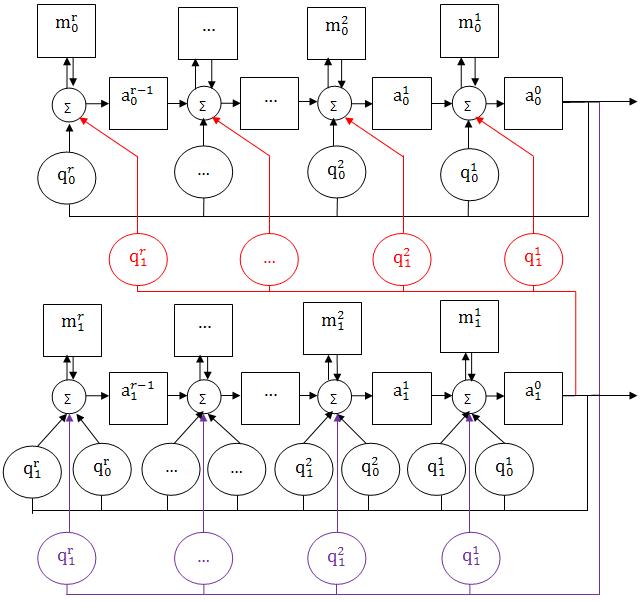}
\end{center}\caption{VFCSR-Q in Galois mode.}
\label{VFCSR-Q-Galois}
\end{figure}

In \cite{allailou2010}, we developed the conception of VFCSR in Galois mode, especially the quadratic case called VFCSR-Q (see Fig \ref{VFCSR-Q-Galois}) and we have presented a new stream cipher design based on a filtered quadratic VFCSR automaton and called F-VFCSR-Q. In the following, we briefly describe VFCSR in Galois mode, analyses basic properties. For more details, we refer to \cite{allailou2010}.
\begin{definition}A Vectorial FCSR in Galois mode over $(\mathbb{F}_{2},P,\mathcal{B})$ of length $r$ with connection coefficients $q_{1},\ldots,q_{r}\in \mathbb{F}_{2}[X]/(P)$ is an automaton whose state is an element $s(t)=(a_{0}(t),\ldots,a_{r-1}(t),m_{1}(t),\ldots,m_{r}(t))$ where $a_{i}(t)\in\mathbb{F}_{2}[X]/(P)$ and $m_{i}(t)\in\mathbb{Z}[X]/(P)$ and whose state change operation is described as follows: 
\begin{itemize}
\item Write elements in the basis $\mathcal{B}$. 
\begin{equation}\label{E6}
\begin{array}{rll}
\forall 0\leq i<r,&a_{i}(t)=a_{0}^{i}(t)+a_{1}^{i}(t)\bar{X}+\ldots+a_{n-1}^{i}(t)\bar{X}^{n-1}&\textrm{ where }a_{j}^{i}(t)\in \left\{0,1\right\},\\
\forall 1\leq i\leq r,&q_{i}=q_{0}^{i}+q_{1}^{i}\bar{X}+\ldots+q_{n-1}^{i}\bar{X}^{n-1}&\textrm{ where }q_{j}^{i}\in \left\{0,1\right\},\\
\forall 1\leq i \leq r,&m_{i}(t)=m_{0}^{i}(t)+m_{1}^{i}(t)\bar{X}+\ldots+m_{n-1}^{i}(t)\bar{X}^{n-1}&\textrm{ where }m_{j}^{i}(t)\in \mathbb{Z}.
\end{array}
\end{equation}
\item Take the canonical lift of the collection of $a_{i}(t)$ and $q_{i}$ in $\mathbb{Z}[X]/(P)$ with respect $\mathcal{B}$. 
\item Compute $\sigma_{i}(t+1)=q_{i+1}a_{0}(t)+a_{i+1}(t)+m_{i+1}(t)$ as a vector in $\mathcal{B}$.  
\item Compute the coordinates of $a_{i}(t+1)$ and $m_{i+1}(t+1)$ wrt $\mathcal{B}$: 
\begin{equation}\label{E4}\begin{array}{rcl}
a_{l}^{i}(t+1)=\sigma_{l}^{i}(t+1) \pmod 2 \textrm{ and }\\
 m_{l}^{i}(t+1)=\frac{1}{2}(\sigma_{l}^{i}(t+1)-a_{l}^{i}(t+1)).\end{array}
\end{equation}
\end{itemize}
$s(0)$ is the initial state, the feedback function is $f(s(t))=s(t+1)$ and the output function is $g(s)=g(x_{0},\ldots,x_{r-1},y_{1},\ldots,y_{r})=(g_{0}(s),\ldots,g_{r-1}(s))=(x_{0},\ldots,x_{r-1})$. The Galois VFCSR generates $r$ vectorial infinite output sequences, for all $0\leq i\leq r-1$:
$$\underline{a}^{i}=(g_{i}( s(0)),g_{i} \circ f(s(0)),g_{i}\circ f^{2}(s(0)),\ldots)=(a_{i}(0),a_{i}(1),a_{i}(2),\ldots).$$ 
\end{definition}
\paragraph{\textbf{Analysis:}}We use the same method as in the Fibonacci case except that we study $r$ output vectorial sequences. Each vectorial output sequence $\underline{a}^{i}$ corresponds to $n$ binary sequences $\underline{a}^{i}_{j}=(a^{i}_{j}(0),a^{i}_{j}(1) \cdots)$. Let $\beta_{j}^{i}=a^{i}_{j}(0) +a^{i}_{j}(1)2+ \cdots$ be the $2$-adic expansion of $\underline{a}_{j}^{i}$ and $\beta$ a $2$-adic vector  associated to a vectorial sequence $\underline{a}$ both of length $nr$.
Simple computations shows that $\beta$ satisfies a linear system with integral coefficients.
This system is represented by an invertible $rn\times rn$ matrix  called \textit{the connection matrix of the Galois VFCSR} also denoted $\mathcal{M}$. $\mathcal{M}$ is equal to the identity matrix minus a matrix with even coefficients.
\begin{equation}\label{E14}
\mathcal{M}=\left(\begin{array}{ccc|ccccccccc}
1-*&\cdots&*&-2&&(0)\\
\vdots&\ddots&\vdots&&-2&\\
*&\cdots&1-*&(0)&&\ddots\\
\hline
*&\cdots&*&1&&(0)\\
\vdots&&\vdots&&1&\\
*&\cdots&*&(0)&&\ddots
\end{array}\right)
\end{equation}
\begin{theorem}Consider a VFCSR in Galois mode over $(\mathbb{F}_{2},P,\mathcal{B})$ of length $r$ with connection integer $q$ and connection matrix $\mathcal{M}$. Then for any sequence $\underline{a}$ generated by this VFCSR, the associated 2-adic vector $\beta $ is in $\frac{1}{|\det \mathcal{M}|}\mathbb{Z}^{nr}$, $|\det \mathcal{M}|$ is odd and $\det(\mathcal{M})=\mathtt{N}(-q)$.
\end{theorem}
VFCSR in Galois mode have the same properties of VFCSRs in Fibonacci mode : periodicity, existence of $l$-sequences etc\ldots Figure \ref{VFCSR-Q-Galois} illustrates VFCSR-Q in Galois mode. We have taken the quadratic case $n=2$ (VFCSR-Q) and the triplet connection in Table \ref{triplet2} to design a cryptographic random generator. For more detail see \cite{allailou2010}.
\begin{table}[!]
\begin{center}
\begin{tabular}{|r|l|}
\hline
$\tilde{q}$=&
3974140296190695420616004753553979604200521434082\\
&082527268932790276172312852637472641991806538949
\\
\hline
$u$=&1993524591318275015328041611344215036460140087963
\\
\hline
$v$=&1993524591318275015328041611344215036460140087860\\
\hline
\end{tabular}
\end{center}
\caption{Example of triplet connection in Galois mode}
\label{triplet2}
\end{table}
\section{Vectorial FCSR in Ring mode}
\begin{figure}[!h]
\begin{center}
\includegraphics[height=8cm,width=7cm]{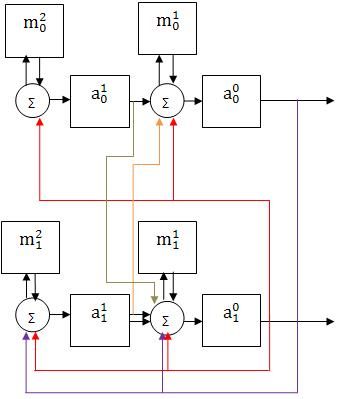}
\end{center}
\caption{Vectorial Feedback with Carry for $\tilde{q}=61$.}
\label{VFCR-Q-Ring-61}
\end{figure}

\begin{definition}[VFCR]A Vectorial Feedback with Carry Register over $(\mathbb{F}_{2},P,\mathcal{B})$ of length $r$ with $r\times r$ transition matrix $T=(t_{i,j})$ and coefficients in $\mathbb{F}_{2}[X]/(P)$ is an automaton whose state is a pair $(a(t),m(t))$ where $a(t)=(a_{0}(t),\ldots,a_{r-1}(t))\in (\mathbb{F}_{2}[X]/(P))^{r}$ and $m(t)=(m_{1}(t),\ldots,m_{r}(t))\in (\mathbb{Z}[X]/(P))^{r}$; and whose operation state change is given by: 
\begin{itemize}
\item Write the collection of $a_{i}(t)$, $m_{i}(t)$ and $t_{i,j}$ in the basis $\mathcal{B}$. 
\item Take the canonical lift of the collection of $a_{i}(t)$ and of $t_{i,j}$ in $\mathbb{Z}[X]/(P)$ with respect $\mathcal{B}$. 
\item Write $a(t)$ and $m(t)$ as vectors of dimension $nr$
\begin{equation}\label{E8}
\begin{array}{rl}
a(t)=&(a_{0}^{0}(t),\ldots,a_{n-1}^{0}(t),\ldots,a_{0}^{r-1}(t),\ldots,a_{n-1}^{r-1}(t))\\
m(t)=&(m_{0}^{1}(t),\ldots,m_{n-1}^{1}(t),\ldots,m_{0}^{r}(t),\ldots,m_{n-1}^{r}(t)).
\end{array}
\end{equation}
\item Replace the multiplication $a_{i}(t)t_{i,j}$ in (\ref{E7}) by the "vectorial" multiplication $\otimes$ in (\ref{E9}) and where $M_{t_{i,j}}$ is the matrix in the canonical basis $\mathcal{B}$ of the linear  transformation defined by the multiplication by $t_{i,j}$. 
\begin{equation}\label{E9}
a_{i}(t)t_{i,j}=(a_{0}^{i}(t),\ldots,a_{n-1}^{i}(t))\otimes M_{t_{i,j}}
\end{equation}
\item From the blocks $M_{t_{i,j}}$, consider the big $rn\times rn$ matrix $\mathcal{T}=(M_{t_{i,j}})_{i,j}$  with coefficients in $\mathbb{Z}$. 
\item Write the addition with $m(t)$ in (\ref{E7}) as a vectorial addition $\oplus$ with the components of $m(t)$ in (\ref{E8}) and compute $a(t)\otimes\mathcal{T}\oplus m(t)$.
\item  Apply $\mod 2$ and $\mathtt{div}2$ componentwise in this equation. 
\end{itemize}
The Ring mode for VFCSR is the case where $t_{i+1,i}=1$ for all $i$.
\end{definition}
\begin{theorem}Consider a VFCR. For all $0\leq i\leq r-1$ and $0\leq j\leq n-1$, the output sequence $(a_{j}^{i}(0),a_{j}^{i}(1),\ldots)$ is associated to a rational number $\frac{p_{i,j}}{\tilde{q}}$ where $\tilde{q}=\det(I_{rn}-2\mathcal{T})$.
\end{theorem}
\paragraph{\textbf{Example 1: FCSR and VFCSR in Fibonacci and Galois mode.}} VFCSR in these both modes can be represented respectively by the following $F$ and $G$ 
\begin{equation}\label{E10}F=\left(\begin{array}{ccccc}
0&\ldots&0&M_{q_{r}}\\
I_{n}&\ldots&0&M_{q_{r-1}}\\
\vdots&\ddots&\vdots&\vdots\\
0&\ldots&I_{n}&M_{q_{1}}
\end{array}\right)\textrm{ and }G=\left(\begin{array}{ccccc}M_{q_{1}}&\ldots&M_{q_{r-1}}&M_{q_{r}}\\
I_{n}&\ldots&0&0\\
\vdots&\ddots&\vdots&\vdots\\
0&\ldots&I_{n}&0
\end{array}\right),
\end{equation}
where $I_{n}$ is the identity matrix of dimension $n$, $0$ is the zero matrix and $M_{q_{i}}$ is the matrix of the linear transformation in $\mathcal{B}$ defined as the multiplication by $q_{i}$. Using linear transformations on lines, we show that $I_{nr}-2F$ can be reduced to a $2  \times 2$ lower triangular block-matrix  with the connection matrix $\mathcal{M}$ in the Fibonacci case and the identity $I_{n(r-1)}$ on the diagonal. The connection matrix of Galois VFCSR in (\ref{E14}) is $I_{rn}-2G^{t}$ where $G^{t}$ is the transpose of $G$. For binary FCSR in Ring mode, $M_{q_{i}}=q_{i}$. 
\paragraph{\textbf{Example 2: VFCR-Q of size $2$.}} a VFCR-Q is a VFCSR over $(\mathbb{F}_{2},X^{2}-X-1,\mathcal{B})$. For $r=2$, the register can be represented by two registers: the main register and the carry register. Each register can be decomposed into two modules of two cells or two carries (see Fig \ref{VFCR-Q-Ring-61}).
The transition matrix $\mathcal{T}$ is of the form (\ref{E12}) and the computations are given by (\ref{E13})
\begin{equation}\label{E12}
\mathcal{T}=\left(\begin{array}{cc}
\begin{array}{cc}
t_{0}^{1,1}&t_{1}^{1,1}\\
t_{1}^{1,1}&t_{0}^{1,1}+t_{1}^{1,1}\\
\end{array}&
\begin{array}{cc}
t_{0}^{1,2}&t_{0}^{1,2}\\
t_{1}^{1,2}&t_{0}^{1,2}+t_{1}^{1,2}\\
\end{array}\\
\begin{array}{cc}
t_{0}^{2,1}&t_{1}^{2,1}\\
t_{1}^{2,1}&t_{0}^{2,1}+t_{1}^{2,1}\\
\end{array}&
\begin{array}{cc}
t_{0}^{2,2}&t_{1}^{2,2}\\
t_{1}^{2,2}&t_{0}^{2,2}+t_{1}^{2,2}\\
\end{array}
\end{array}
\right)
\end{equation}
\begin{equation}\label{E13}
(a_{0}^{0}(t),a_{1}^{0}(t),a_{0}^{1}(t),a_{1}^{1}(t))\otimes \mathcal{T}\oplus (m_{0}^{1}(t),m_{1}^{1}(t),m_{0}^{2}(t),m_{1}^{2}(t)).
\end{equation}
We can built $2^{nr^{2}}$ distinct VFCRs over $\mathbb{F}_{2^{n}}$ of size $r$. Among all binary FCR of size $4$, the maximal period is $60$ and there is a VFCR-Q of size $2$ generating a sequence with this period (see Table \ref{comparaison}). For example, with the transition matrix $T$ bellow which correponds to the transition matrix $\mathcal{T}$ (\ref{E15}), we can generate two vectorial sequences with period $\mathtt{ord}_{\tilde{q}}(2)=60$ where $\tilde{q}=|\det(I-2\mathcal{T}_{0})|=61$. We have loading initial state $(a_{0},a_{1},m_{1},m_{2})=(1+\bar{X},1,0,\bar{X})$ and output the sequence of Table \ref{example}.
\begin{equation}\label{E15}
\begin{array}{cc}
T=\left(
\begin{array}{cc}
\overline{X}&\overline{X}\\
1+\overline{X}&0
\end{array}
\right)&
\quad , \quad \mathcal{T}=
\left(
\begin{array}{cccc}
0&1&0&1\\
1&1&1&1\\
1&1&0&0\\
1&2&0&0
\end{array}
\right)
\end{array}
\end{equation}
\begin{table}[h]
\begin{center}
\begin{tabular}{|l|c|l|l|}
\hline
Registers&differents&values &maximal period\\
&models&$\tilde{q}=|\det(I-2T)|$& $\mathtt{ord}_{\tilde{q}}(2)=\tilde{q}-1$\\
\hline
binary FCR of size 2& $2^{4}$&1,3,5&2,4\\
\hline
binary FCR of size 4& $2^{16}$& 1,3,5,7,9,$\cdots$,59,61,63,&2,4,10,12,18,\\
&&69,75,77,81,87,91,99,135&28,36,52,58,60\\
\hline
VFCSR-Q in Fib. &$2^{4}$&1,5,9,11,19,25,29&4,10,18,28\\
and Gal. of size 2&&,31,41&\\
\hline
VFCR-Q&$2^{8}$&1,5,9,11,19,25,29,&4,10,18,28,60\\
of size 2&&31,41,45,49,55,61,99&\\
\hline
\end{tabular}
\end{center}
\caption{Comparaison of maximal periods of FCR of size $2,4$ and VFCR-Q of size $2$.}
\label{comparaison}
\end{table}
\begin{table}[!h]
\begin{center}
\begin{tabular}{|c|l|}
\hline
$a_{0}^{0}$&1     0     0     0     1     1     1     0     1     0     0     1     0     0     1     1     0     0     0     0     1     0     0  0     0     0     1     1     0     1     0     1     1     1     0     0     0     1     0     1     1     0     1     1     0     0\\
\hline
$a_{1}^{0}$&     1     1     1     0     1     1     1     1     1     0     0     1     0     1     0     0     0     1     1     1     0     1     0  0     1     0     0     1     1     0     0     0     0     1     0     0     0     0     0     1     1     0     1     0     1     1\\
\hline
$a_{0}^{1}$&     1     1     1     1     0     1     1     1     1     1     0     0     1     0     1     0     0     0     1     1     1     0     1  0     0     1     0     0     1     1     0     0     0     0     1     0     0     0     0     0     1     1     0     1     0     1\\
\hline
$a_{1}^{1}$&     0     1     0     0     1     0     1     1     0     1     1     0     0     1     1     1     1     0     1     1     1     1     1  0     0     1     0     1     0     0     0     1     1     1     0     1     0     0     1     0     0     1     1     0     0     0\\
\hline

\hline
 $a_{0}^{0}$&1     1     1     1     0     1     1     1     1     1     0     0     1     0     1     0     0     0     1     1     1     0     1     0     0     1     0     0     1     1     0     0     0     0     1     0     0     0     0     0     1     1     0     1     0     1\\
 \hline
$a_{1}^{0}$&     1     0     0     0     1     0     1     1     0     1     1     0     0     1     1     1     1     0     1     1     1     1     1       0     0     1     0     1     0     0     0     1     1     1     0     1     0     0     1     0     0     1     1     0     0     0\\
\hline
 $a_{0}^{1}$&     1     1     0     0     0     1     0     1     1     0     1     1     0     0     1     1     1     1     0     1     1     1     1      1     0     0     1     0     1     0     0     0     1     1     1     0     1     0     0     1     0     0     1     1     0     0\\
 \hline
  $a_{1}^{1}$    &      0     1     0     0     0     0     0     1     1     0     1     0     1     1     1     0     0     0     1     0     1     1     0        1     1     0     0     1     1     1     1     0     1     1     1     1     1     0     0     1     0     1     0     0     0     1\\
 \hline
\end{tabular} 
\end{center}
\caption{Example of VFCR-Q sequence of period 60.}
\label{example}
\end{table}
\section{Vectorial memory requirements}
\begin{figure}[!h] 
\begin{center}
\includegraphics[]{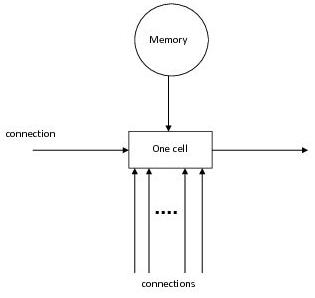}
\end{center}\caption{Representation of cell and its connections.}
\label{connection}
\end{figure}

It's important to describe the memory behavior when the register runs. Concretely, each cell has a determined number of connections (with other cells of the main register) over the connection to the memory cell corresponding (see figure \ref{connection}). It exists a range of values stable for the memory.
\begin{theorem}Consider a VFCR with vectorial transition matrix $\mathcal{T}$. Call $\mathcal{C}^{i}_{j}$ the $(in+j)$-th column of $\mathcal{T}$ and $w_{j}^{i}$ the sum of its coefficients. Let $(a(t),m(t))$ the state of the $t$-th step of the register. The coordinates of the next state are given by the following recursive relation:
$a(t).\mathcal{C}^{i}_{j}+m_{j}^{i}(t)=a_{j}^{i}(t+1)+2m_{j}^{i}(t+1).$
If $m_{j}^{i}(t)\in [0,w_{j}^{i}[$, then $m_{j}^{i}(t+1)\in [0,w_{j}^{i}[$.
\end{theorem}
For example, with the transition matrix $\mathcal{T}$ (\ref{E15}) and the initial state $(1+\bar{X},1,0,\bar{X})$, we obtain these following values for the memories:
\begin{table}[h]
\begin{center}
\begin{tabular}{|c|c|}
\hline
$m_{0}^{0}$&0     1     2     2     1     1     1     2     2     2     2     1     1     1     1     1     1     1     1     2     2     2     2 2     1     1     1     0     1     1     1     0     0     0     1     1     1     0     0     0     0     1     1     1     1      $\cdots$\\
\hline
$m_{1}^{0}$&     0     1     2     2     1     2     2     3     3     3     3     2     2     1     2     3     3     2     1     2     3     3     3 3     1     1     2     1     2     2     2     1     2     2     3     2     2     1     1     1     1     2     2     3     2      $\cdots$\\
     \hline
 $m_{0}^{1}$&    0     0     0     0     0     0     0     0     0     0     0     0     0     0     0     0     0     0     0     0     0     0     0    
     0     0     0     0     0     0     0     0     0     0     0     0     0     0     0     0     0     0     0     0     0     0      $\cdots$\\
     \hline
$m_{1}^{1}$&     1     1     1     1     0     1     1     1     1     1     0     0     1     0     0     0     0     0     0     0     0     0     0 0     0     0     0     0     1     1     1     0     0     0     1     0     0     0     0     0     1     1     0     1     1      $\cdots$\\
 \hline
\end{tabular} 
\end{center}
\caption{Memory values.}
\label{example2}
\end{table}
For example, with the vectorial transition matrix $\mathcal{T}$ (\ref{E15}) and the initial state $(1+\bar{X},1,0,\bar{X})$, we obtain the memory values of the Table \ref{example2} and we can see that $m_{0}^{0}$ returns and remains in the interval $[0,w_{0}^{0}[$, $m_{1}^{0}$ in $[0,w_{1}^{0}[$, $m_{0}^{1}$ in $[0,w_{0}^{1}[$ and $m_{1}^{1}$ in $[0,w_{1}^{1}[$ where $w_{0}^{0}=3$, $w_{1}^{0}=5$, $w_{0}^{1}=1$ and $w_{1}^{1}=2$.
\section{Conclusion}
We extended the notion of VFCSR to the notion of VFCR which are defined by an arbitrary transition matrix. This allows to vary the model register playing with the connections and to construct FCR over $\mathbb{F}_{2^{n}}$. On the other hand, VFCR structure allowed to extract $n$ bytes every time the generator is clocked, it is more efficient than the classical FCR. Moreover, we can obtain maximal periods greather than those of the classical models called Fibonacci, Galois or Ring.
\bibliographystyle{amsalpha}

\end{document}